\documentclass[preprint2]{aastex}
\shortauthors{Hodge et al.}
\shorttitle{C32, A Young Star Cluster in IC 1613}

\begin{document}

\title{C32, A Young Star Cluster in IC 1613}

\author{Ted K. Wyder, Paul W. Hodge} 
\affil{Astronomy Department,
University of Washington, Box 351580, Seattle, WA 98195-1580;
wyder@astro.washington.edu, hodge@astro.washington.edu}

\and

\author{Andrew Cole}
\affil{Department of Physics \& Astronomy, 640 Lederle Grad Research Center, University of Massachusetts, Amherst, MA 01003; cole@condor.astro.umass.edu}

\begin{abstract}

The Local Group irregular galaxy IC 1613 has remained an enigma for
many years because of its apparent lack of star clusters. We report
the successful search for clusters among several of the candidate
objects identified many years ago on photographic plates. We have used
a single HST WFPC2 pointing and a series of  images obtained with the
WIYN telescope under exceptional seeing conditions, examining a total
of 23 of the previously published candidates. All but six of these
objects were found to be either asterisms or background galaxies. Five
of the six remaining candidates possibly are  small, sparse clusters
and the sixth, C32, is an obvious cluster. It is a compact, young
object, with an age of less than 10 million years and a total absolute
magnitude of M$_V = -5.78\pm0.16$ within a radius of 13 pc.

\end{abstract}


\section{Introduction}
\citet{ba63} remarked on the fact, which he obviously considered
remarkable, that the irregular galaxy IC 1613, a member of the Local Group
that he had studied quite intensively, appeared to be devoid of star
clusters. This fact was again discussed by \citet{va79}, who
compared IC 1613 with the SMC, pointing out that the SMC's many rich
clusters are so conspicuous that, if IC 1613 has any, they should show up
clearly on available plate material. \citet{ho78} had searched photographic
plates taken with the Palomar Observatory's 5m and the Lick Observatory's
3m telescopes and published a list of 43 possible candidates for clusters,
all of which were very inconspicuous, none having more than 6 resolved
stars on the best images. Clearly, these objects were not comparable to the
rich star clusters of the Magellanic Clouds, but he thought that at least
some of them might be similar to small Galactic open clusters like the
Pleiades. More recently \citet{fr88}, in her CCD-based study of the
color-magnitude diagram (CMD) of a portion of IC 1613, noted that she could
not identify those candidate clusters that should have been visible on her
images. On the other hand, \citet{ge99} observed  a field
centered on the most active area of star formation (east of our present
field), identified 12 of the earlier cluster candidates and measured
integrated UBV magnitudes for eight, six of which they found to have
colors indicating ages of less than 10 million years. They also identified
two new small, young clusters. They searched for massive clusters, either
young or old globular clusters, but found none. For a related discsussion
of the young massive cluster populations in galaxies and their
correlations with integrated galaxy properties see \citet{la00}.

Thus it has been  clear that IC 1613 must be poor in rich star
clusters compared with the Magellanic Clouds, a fact which van den
Bergh attributed to the lack of a history of strong shocks. But it has
not been clear whether or not IC 1613 is lacking in a significant
number of normal open clusters, as most of the published candidate
clusters have not been examined on anything but the original
photographic plates. This paper reports the results of a study of some
of these candidates using HST and WIYN\footnote{The WIYN Observatory
is a joint facility of the University of Wisconsin-Madison, Indiana
University, Yale University, and the National Optical Astronomy
Observatories.} telescope images.

\section{Observational Material}

For our cluster search we have used multi-color images of IC 1613 that
were obtained to study the stellar populations in IC 1613 , which have
been fully discussed elsewhere \citep{co99}. One set of images is a
single pointing of HST centered near the geometrical center of IC 1613
at RA = $1^h$ $04^m$ $48.7^s$, declination = $+02\arcdeg$ 07$\arcmin$
06.2$\arcsec$ (J2000). WFPC2 images were taken with the F439W, F555W
and F814W filters. All are available from the HST Data Archive (PID
6865). The other set was obtained by  the WIYN telescope at Kitt Peak
towards the same position using a CCD detector and B, V and I filters,
taken under 0.6 arcsec seeing conditions.

Both sets of images are centered on the HI "hole"
\citep{sk87,la89,we96} at the center of the galaxy and do not include
the most active star-forming regions, which are concentrated primarily
to the east of these images \citep{ho78,ho90,pr90}. The WIYN
images are 6.8 arc minutes on a side, providing an area of 46.24
square arc minutes, while the entire area of IC 1613, to the V = 25
magnitudes per square arcsec isophote, is 283 square arc minutes. Thus
the region surveyed here is about 16\% of the main body of the galaxy.

\section{Survey Results}
The HST images included only two of the previously-identified cluster
candidates, nos. 22 and 33, both of which had been considered faint and
doubtful. The HST images revealed both to be small asterisms, with no
indication of the presence of any physical clustering of stars (Fig. 1).

The WIYN images covered an additional 21 candidate clusters. One of
these, cluster C32, has the appearance of a true, small but bright
cluster (Fig. 2). Five other candidates are possible examples of small,
very sparse clusters, for which the WIYN images are not definitive. The
others are either small asterisms or background galaxies with
foreground stars superimposed, making them appear to be star clusters
when imaged with insufficient resolution.  Table 1 lists the candidates
examined and their disposition.

\section{Cluster C 32}
Cluster C32  (Fig. 2) is a small cluster located near the center of OB
association No. 5 \citep{ho78}. The OB association is involved with HII
region S2 \citep{sa71}, which is cataloged as HLG 13 \citep{ho90}.
The cluster is thus located in a small area of current star formation.
The entire WIYN image is shown in Figure 3 with the position of C32 indicated.

Figure 4 shows the radial distribution of stars in the cluster and its
surroundings. The half-light diameter of the cluster is only of the
order of 2.0 arcsecs. One arcsec at IC 1613's distance \citep{le93}
corresponds to 3.5 parsecs, implying that cluster C32 is approximately
7 parsecs across, making it similar in size to many small young open
clusters in the Milky Way Galaxy.

Photometry of the stars in the WIYN B and V images was extracted using
the PSF-fitting software DAOPHOTII and ALLSTAR \citep{st94}. The
magnitude zeropoints were determined via comparison with the B and V
photometry of \citet{fr88}. We first determined an astrometric
solution for the WIYN data using the Digitized Sky Survey images
available from the Space Telescope Science Institute. We then matched
together the positions of relatively bright stars measured by Freedman
in her Field 1 that appear in our WIYN images. The average offset
between our instrumental psf-fit magnitudes and Freedman's photometry
was determined using a total of 24 and 25 stars in B and V,
respectively. The rms scatter for the sample of comparison stars was
0.05 magnitudes in both filters.

Figure 5 is a color-magnitude diagram that shows the positions of the
few well-separated stars of C32 detected in both B and V superimposed
on the CMD of the entire field of the WIYN images.  Clearly the
cluster is very young, with a main sequence that extends to M$_V =
-5$. There are too few stars to establish a definitive age from the
CMD, but the cluster is clearly younger than about 10 million years, a
conclusion that agrees with its position in a young OB association
that is enveloped in an HII region.

To measure the integrated magnitude and color of C32, we first
calculated the background as the mean flux in an annular region
between $6\arcsec$ and $8\arcsec$ from the center of C32. This value
was then subtracted from the flux within a circular aperture of radius
13 pc. The error in the flux was determined by randomly choosing 100
aperture centers within a $1.6\arcmin \times 1.6\arcmin$ area
surrounding C32. At each of these positions, we calculated the flux
and background using the same aperture and sky annulus as used for the
cluster. The resulting flux values had an average of zero and a
standard deviation that we assume as the probable error in the
aperture magnitude of C32. Assuming a foreground reddening of
E(B$-$V)$=0.025$ \citep{sc98}, the integrated color and magnitude of
C32 within a radius of 13 pc are M$_V=-5.78\pm0.16$ and
(B$-$V)$_0=-0.19\pm0.08$. These values of color and absolute magnitude
are fully consistent with those of small young star clusters in other
galaxies, such as M31 \citep{ho87}.

It is interesting to make a quantitative comparison of the cluster
density in IC 1613 with that in other galaxies. We have counted the
number of star clusters found in the LMC on small-scale plate surveys
(approximately equivalent in physical scale and limiting absolute
magnitude to the WYIN survey of IC 1613) in three areas each equal in
physical size to the WYIN IC 1613 field and located at a corresponding
optical surface brightness. While we have found only one certain star
cluster in the IC 1613 field, the LMC fields averaged $81\pm9$
clusters. Details of this comparison, as well as comparisons with
other galaxies, are given in \citet{ho00}.

\section{Conclusions}

We have surveyed the central area of IC 1613 for star clusters. There are
no globular clusters or massive young clusters present, but there are six
very sparse groupings that may be open clusters like some of the smaller
ones in our Galaxy. One of them, previously cataloged as C32, is a very
young cluster embedded in an HII region and surrounded by an OB
association.  When the uncertain nature of many of the cluster candidates,
their very small sizes and their implied small stellar populations are
taken into account, it is clear that IC 1613 is a cluster-poor galaxy.

\acknowledgements{We are grateful to NASA's Space Telescope Science
Institute for partial support of this research through grant AR-08362
to the University of Washington. The original observations were made
with the NASA/ESA Hubble Space Telescope, obtained at the Space
Telescope Science Institute, which is operated by the Association of
Universities for Research in Astronomy.  Inc., under NASA contract NAS
5-26555, and with the WIYN Telescope at the Kitt Peak National
Observatory.}

\newpage
\centerline{Figure Captions}

\figcaption{The WFPC2 V image of the central area of IC 1613 with the
positions of cluster candidates C22 and C33 indicated.}

\figcaption{Cluster C32, from the WIYN image in V.}

\figcaption{The entire WIYN V image of the central $6.8\arcmin \times
6.8\arcmin$ area of IC 1613 with North up and East to the left. The
position of cluster C32 is indicated.}

\figcaption{The radial profile of cluster C32 (large dots and solid
line), determined from the WIYN V image. The horizontal line indicates
the average value in an annular region between $6\arcsec$ and
$8\arcsec$ from the center of C32.  The dashed line is the profile of
an isolated star in the image scaled to the peak brightness of C32.}

\figcaption{The color-magnitude diagram of cluster C32 (large dots)
and the field of the entire WIYN image. The error bars represent
the average errors in color and magnitude as a function
of magnitude.}

\begin{deluxetable}{cc}
\tablecaption{The Nature of the Cluster Candidates}
\tablewidth{0pt}
\tablehead{
	\colhead{Cluster No. from Hodge (1978)} & \colhead{Description}
}
\startdata
C 9	& asterism \\
C 19	& probable asterism \\
C 20	& asterism \\
C 21	& galaxy \\
C 22	& asterism \\
C 23	& possible red cluster \\
C 24	& asterism \\
C 25	& asterism \\
C 26	& possible cluster \\
C 27	& possible cluster \\
C 28	& possible cluster \\
C 31	& asterism \\
C 32	& cluster \\
C 33	& asterism \\
C 34	& galaxy pair? \\
C 35	& asterism \\
C 36 	& asterism \\
C 37	& asterism \\
C 38	& asterism \\
C 39	& asterism or galaxy \\
C 41	& asterism \\
C 42	& possible loose cluster \\
C 43 	& probable asterism \\
\enddata
\end{deluxetable}


\begin{thebibliography}{}

\bibitem[Baade(1963)]{ba63} Baade, W. 1963, Evolution of Stars and Galaxies (Cambridge: Harvard U. Press)
\bibitem[Cole et al.(1999)]{co99} Cole, A. A. et al. 1999, \aj, 118, 1657
\bibitem[Freedman(1988)]{fr88} Freedman, W. 1988, \aj, 96, 1248
\bibitem[Georgiev et al.(1999)]{ge99} Georgiev, L., Borissova, J., Rosado, M., Kurtev, R., Ivanov, G., \& Koenigsberger, G. 1999, \aaps, 134, 21
\bibitem[Hodge(1978)]{ho78} Hodge, P. W. 1978, \apjs, 37, 145
\bibitem[Hodge(2000)]{ho00} Hodge, P. W. 2000, in preparation
\bibitem[Hodge et al.(1990)]{ho90} Hodge, P. W., Lee, M.-G., \& Gurwell, M. 1990, \pasp, 102, 1245 
\bibitem[Hodge et al.(1987)]{ho87} Hodge, P., Mateo, M., Lee, M. G., \& Geisler, D. 1987, \pasp, 99, 173
\bibitem[Lake \& Skillman(1989)]{la89} Lake, G. R. \& Skillman,  E. D. 1989, \aj, 98, 1274
\bibitem[Larsen \& Richtler(2000)]{la00} Larsen, S. S. \& Richtler, T. 2000, \aap, in press
\bibitem[Lee et al.(1993)]{le93} Lee, M.-G., Freedman, W. L., \& Madore, B. F.  1993, \apj, 417, 553
\bibitem[Price et al.(1990)]{pr90} Price, J. Mason, S. F., \& Gullixson, C. A. 1990, \aj, 100, 420 
\bibitem[Sandage(1971)]{sa71} Sandage, A. R. 1971, \apj, 166, 13 
\bibitem[Schlegel et al.(1998)]{sc98} Schlegel, D. J., Finkbeiner, D. P., \& Davis, M. 1998, \apj, 500, 525 
\bibitem[Skillman(1987)]{sk87} Skillman, E. D. 1987, in Star Formation in Galaxies (Washington:NASA), p.263 
\bibitem[Stetson(1994)]{st94} Stetson, P. B. 1994, \pasp, 106, 250
\bibitem[van den Bergh(1979)]{va79} van den Bergh, S.  1979, \apj, 230, 95
\bibitem[Westpfahl et al.(1996)]{we96} Westpfahl, D., Wilcots, E. Graham, M., \& Olsen, K. 1996, \baas, 28 891

\end{thebibliography}
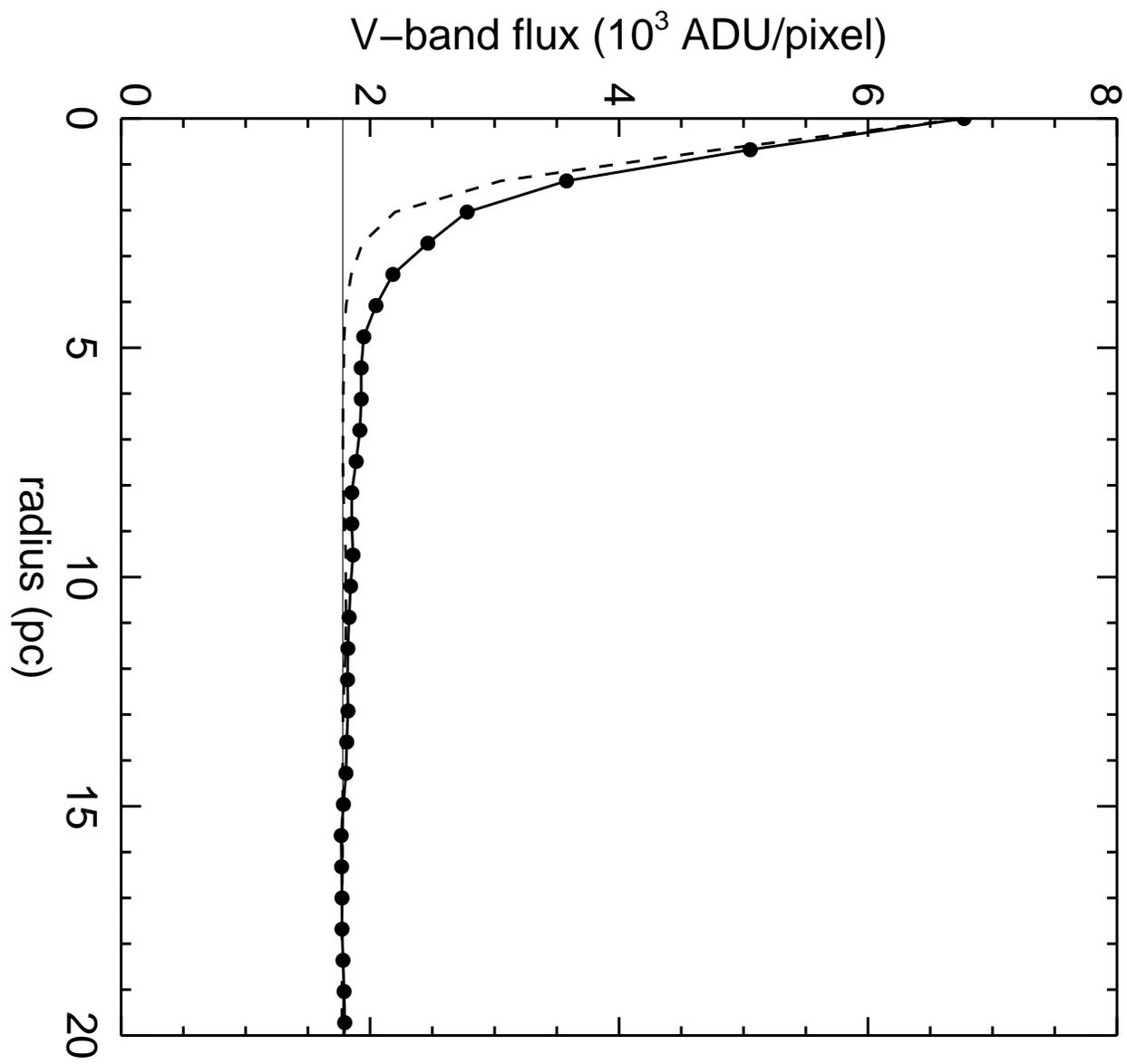
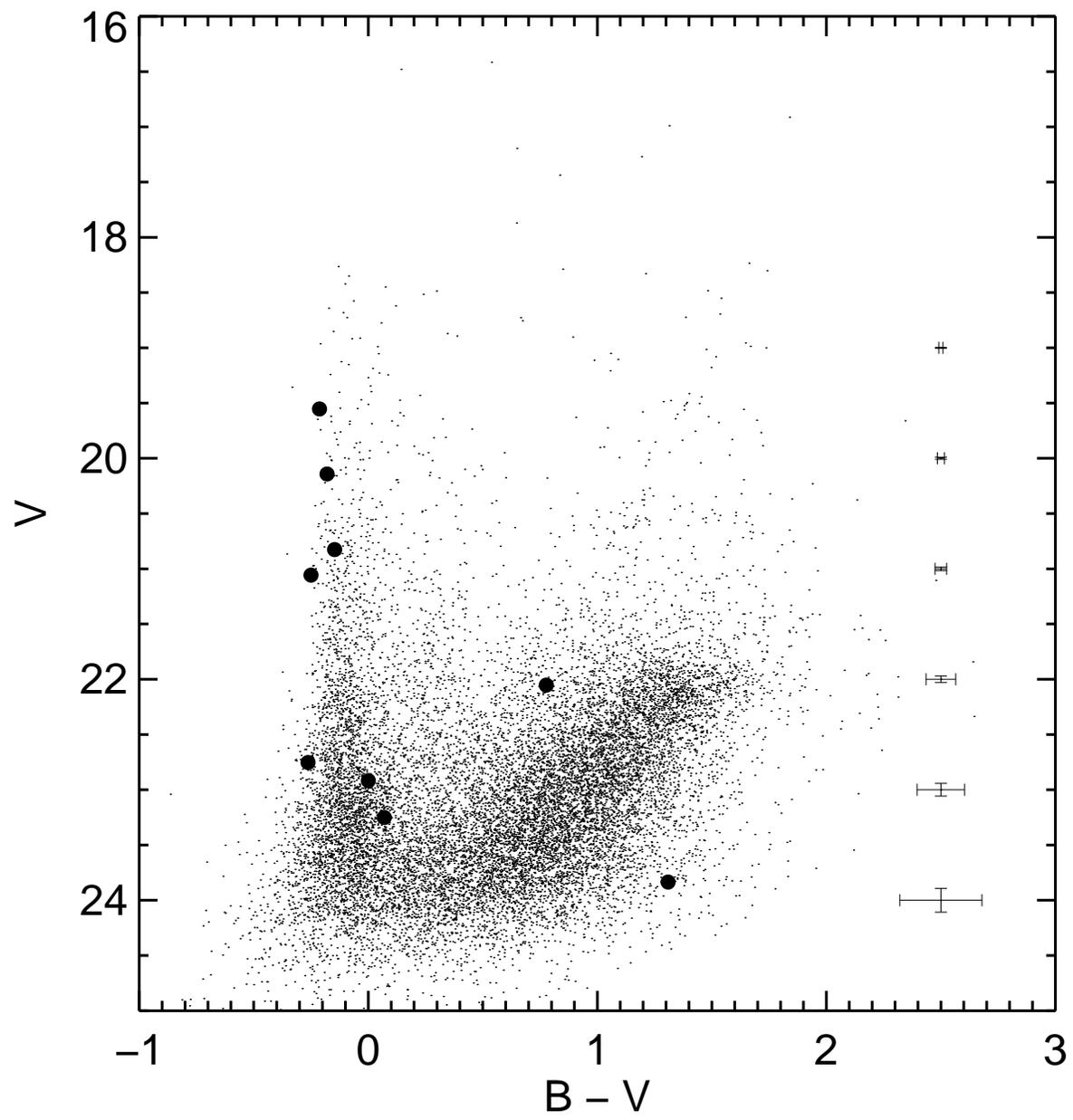
\end{document}